\documentclass[aps,pre,amsmath,amssymb,twocolumn,groupedaddress,showpacs]{revtex4}
\usepackage{graphicx}
\usepackage[utf8]{inputenc}

\usepackage{epsf}

\newcommand{\be}{\begin{equation}}
\newcommand{\ee}{\end{equation}}
\newcommand{\ba}{\begin{eqnarray}}
\newcommand{\ea}{\end{eqnarray}}
\newcommand{\baa}{\begin{eqnarray}}
\newcommand{\eaa}{\end{eqnarray}}
\newcommand{\ed}{\end{document}}

\newcommand{\re}[1]{(\ref{#1})}

\renewcommand{\baselinestretch}{1.2}
\date{\today}
\begin{document}
\title{Ideal quantum gas in expanding cavity: \\
nature of non-adiabatic force}
\author{K.Nakamura$^{(1,3)}$, S.K.Avazbaev$^{(2)}$, Z.A. Sobirov$^{(2)}$, D.U. Matrasulov$^{(2)}$, T. Monnai$^{(3)}$}
\affiliation{$^{(1)}$Faculty of Physics, National University of Uzbekistan,
Vuzgorodok, Tashkent 100174, Uzbekistan \\
$^{(2)}$Turin Polytechnic University in Tashkent,
 17 Niyazov str. (Small Ring),  Tashkent 100094, Uzbekistan \\
$^{(3)}$Department of Applied Physics, Osaka City University, Osaka 558-8585, Japan }

\begin{abstract}
We consider a quantum gas of non-interacting  particles confined in the expanding cavity,
and investigate the nature of the non-adiabatic force which is generated from the gas and
acts on the cavity wall.  Firstly, with use of the time-dependent canonical transformation which transforms the expanding cavity to the
non-expanding one, we can define the force
operator. Secondly, applying the perturbative theory which works when the cavity wall
begins to move at time origin,  we find that the non-adiabatic force is quadratic in the wall
velocity and thereby does not break the time-reversal symmetry, in contrast with the general belief.  Finally, using an assembly
of the transitionless quantum states, we obtain the  nonadiabatic force exactly. The exact result
justifies the validity of both the definition of force operator and the issue of the perturbative theory.
The mysterious mechanism of nonadiabatic transition with use of transitionless quantum states
is also  explained. The study is done on both cases of  the hard-wall and soft-wall
confinement with the time-dependent confining length.

\end{abstract}
\maketitle

\section{ Introduction}

The nonadiabatic transition in the quantum gas of non-interacting  particles confined in an expanding microscopic cavity is the origin of the nonadiabatic force acting on  the cavity wall. Let's consider non-interacting Fermi particles confined in a billiard with a moving wall. The
wall receives  the forces from Fermi particles
in the billiard. Under the condition that whole system consisting of Fermi particles and a moving wall keeps the energy conservation, the work done on the wall by the force is supplied by the excess energy due to the energy loss of Fermi particles
showing the non-adiabatic transition. In this way one can conceive both the adiabatic
and nonadiabatic forces. In the adiabatic limit, the adiabatic force due to the quantal gas on the cavity wall is proportional to the derivative of the confining energy with respect to the cavity size. What is a characteristic feature of the nonadiabatic force when the cavity wall is moving?
The general belief is that this force should be linear in the wall velocity, breaking the time reversal symmetry. In fact, in
compound systems like molecules where two kind of coordinates with different time scales coexist, the Born-Oppenheimer approximation
leads to both the adiabatic and nonadiabatic forces acting on the degree of freedom characterized by the slow time scale, and the
latter force is linear in the velocity of the slow degree of freedom and is called an irreversible or frictional force \cite{robbin,sinc,jarz}. In the case of the expanding cavity, however,  the Hilbert space as well as the domain of Hamiltonian is time-dependent because of the time-dependent  length scale of the cavity confining particles, which requires a deeper insight into the nature of the nonadiabatic force.

The investigation of the above subject was started by Hill and Wheeler in 1952 \cite{hill52} in the context of nuclear physics.
The nature of the nonadiabatic force on the cavity wall were  intensively studied by Blocki {\it et al.} \cite{bloc}.
Wilkinson developed the extensive study on this subject \cite{wil87,wil95} in the context of energy diffusion and of random matrix theory assimilating the chaotic motion of particles inside the cavity, which was followed by other groups \cite{mizu,bulgac,coh, aus}.
Most of these works regarded the force due to the quantal gas as conjugate to a time-dependent wall coordinate.
However, the definition of the force operator is not clear at all in the case of the hard wall.  In fact, one cannot define the force operator
by using a given Hamiltonian for the billiard with a moving boundary. Further, without the valid definition of the force operator,
essential results so far would be highly questionable.

\begin{figure}[htb]
\centerline{\includegraphics[width=\columnwidth]{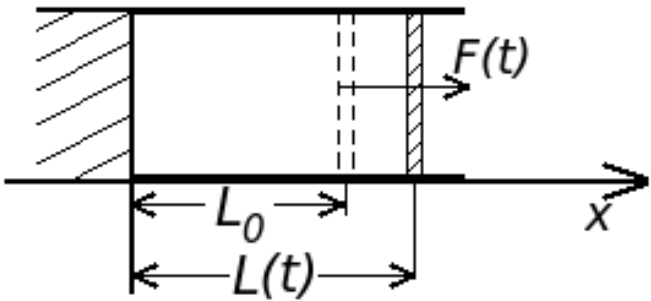}}
\caption{Moving wall confining the quantal gas. $L(t)$ is the time-dependent size of the cavity, and $F(t)$ stands for the sum of adiabatic and non-adiabatic forces.}
\label{mov-wall}
\end{figure}

In this paper, we introduce the force operator with use of the time-dependent canonical transformation
which transforms the expanding cavity to the non-expanding one.
Applying the perturbative and exact theories, we evaluate the non-adiabatic
force  whose nature thoroughly differs from the conventional frictional force. For comparison
we shall also investigate the case of the soft-wall confinement with the time dependent confining length.

In Section \ref{f-operator}, we construct the force operator acting on the moving wall in an unambiguous way.
We consider the case that the cavity wall suddenly begins to move at time origin (see Fig.\ref{mov-wall}).
In Section \ref{perturb}, within a framework of von Neumann equation for the density operator,
we apply a perturbative theory to obtain the expectation of the nonadiabatic force.
In Section \ref{exact}, with use of the transitionless basis functions \cite{mak91,mak92, raz91},
we exactly
evaluate the energy expectation to see the nature of the nonadiabatic force,
and justify the validity of both the definition of force operator and the issue
of the perturbative theory. In Section \ref{soft}
we investigate the case of soft-wall confinement by treating a tunable harmonic trap.
Section \ref{concl} is devoted to summary and discussions. In Appendix \ref{nonconst_vel} we investigate the exactly-solvable case
when the expansion rate of the cavity is time-dependent, to see the universality of the assertion of the text. Appendix \ref{IJ} treats technical details of the integrals used in Section \ref{exact}.

\section{ Force operator} \label{f-operator}
When a given cavity in 1 dimension has a size $L$ and its wall is fixed, the force on the wall due to the quantal gas inside the cavity is defined
by $F=-\frac{\partial E_n}{\partial L}$, which, with use of eigenvalues $E_n=\frac{\hbar ^2 n^2}{2m L^2}$,
gives rise to $F=\frac{\hbar ^2 n^2}{m L^3}$. And the contribution from all particles is expressed as
$F=\sum_{n=1}^{\infty}\frac{\hbar ^2 n^2}{m L^3}f(E_n)$ where $f(E_n)$ stands for the Fermi distribution function.
At zero temperature, $f(E_n)=1$ for $1 \le n \le N$ and $f(E_n)=0$ otherwise. This force is called the adiabatic force.
When the wall will move, the wall receive the extra force depending on its velocity, which comes from the nonadiabatic transition occuring in the quantal gas. However, the definition of the force operator is far from obvious in the case of a moving hard wall. Below we shall define the force operator in two ways.

\subsection{Classical force and quantization}
As a first step to provide the force operator, we show a kinetic evaluation of
the force due to the classical ideal gas.
Then the force is expressed as a dynamical quantity, and is thus quantized
straightforwardly.

Suppose that $N$ mutually noninteracting particles with a common mass $m$ are
confined in a 3-dimensional (3-d) box whose 3 edges have a common length $L$.
Along each of 3 coordinates, one wall is fixed at the origin, for instance, at
$x=0$, and another one is initially located, for instance, at $x=L$ and begins
to move with a constant velocity $(\dot{L})$.
The velocity of the wall $\dot{L}$ is assumed to be slow compared with the
mean square velocity of the particles $\sqrt{\langle v_i^2\rangle}$.
In the course of time evolution, the particles eventually become uniformly
distributed in the box.
The  time $t$ for a particle at a position $x=X_i$ running to right with $v_i(>0)$ (to left with $-v_i$)
to reach the moving wall is given by $t=\frac{L\mp X_i}{v_i-\dot{L}}$ and its average is
\begin{equation}
t=\frac{L}{v_i-\dot{L}}.
\end{equation}
The average time for  a particle to come back to the initial position is given by $2t$.
On the other hand, the momentum change at each collision with the moving wall is given by
\begin{equation}
\Delta p_i=m(v_i-\dot{L})-m(-v_i-\dot{L})=2mv_i.
\end{equation}

Since the collision rate is given as the inverse of $2t$, the force acting on the wall is given by
\begin{eqnarray}
&&\sum_{i=1}^N \frac{v_i-\dot{L}}{2L}\Delta p_i \nonumber \\
&=&\sum_{i=1}^N \left(\frac{v_i}{L}m v_i -\frac{\dot{L}}{L}m v_i\right).\label
{classicalforce}
\end{eqnarray}
The first and second terms on the second line are the adiabatic and non-adiabatic forces, respectively.
These forces are rewritten as
\begin{eqnarray}
&&\sum_{i=1}^N \frac{m v_i^2}{L}=\langle \frac{p^2}{mL} \rangle ,
\end{eqnarray}
and
\begin{eqnarray}
-\sum_{i=1}^N \frac{\dot{L}}{2L^2}(L mv_i + mv_i L) =-\langle \frac{\dot{L}}
{2L^2} (x p +p x)\rangle,
\end{eqnarray}
where $p$ is the momentum.
For each particle, the bracket evaluates the value of the position $x$ and
momentum $p$ at the instance of collision.

Then the non-adiabatic force is given as
\begin{equation}
\label{phenom}
F_{non-ad}=-\frac{\dot{L}}{2L^2} (x p + p x).
\end{equation}
As a dynamical quantity, $F_{non-ad}$ is invariant under the time
reversal operation,
since both the expansion rate $\dot{L}$ and momentum $p$ change their signs.

Let's quantize the non-adiabatic force obtained above.
The force operator should satisfy the following conditions.
\begin{itemize}
\item[a)]In the classical limit, $\hat{F}_{non-ad}$ agrees with $-\frac{\dot{L}}
{2L^2} (x p +p x)$.
\item[b)]$\hat{F}_{non-ad}$ should be Hermitian.
\item[c)]$\hat{F}_{non-ad}$ does not depend on the particle statistics (boson or
fermion).
\end{itemize}
Consequently the force operator should be
\begin{equation}
\label{phenom}
\hat{F}_{non-ad}=-\frac{\dot{L}}{2L^2} (\hat{x}\hat{p} +\hat{p}\hat{x}).
\end{equation}

The idea above is based on the phenomenological argument with use of a 3-d
box, but suggesting a promising expression of the force operator.
Below we shall provide a rigorous definition of the force operator.

\subsection{ Rigorous definition of force operator via time-dependent canonical transformation}
The original Hamiltonian $H$ for the billiard with a time-dependent cavity size $L(t)$ is given by (in unit of $\hbar^2/m=1$)
\begin{align}
H=-\frac{1}{2}\frac{\partial^2}{\partial x^2}.
\label{orig-H}
\end{align}
We now see the expectation of $H$ as given by
\begin{equation}
\langle \psi |H|\psi \rangle,
\end{equation}
where $|\psi  \rangle$ is a solution of the time-dependent Schr\"odinger equation
\begin{align}
i\hbar \frac{\partial }{\partial t}\psi(x,t) =H\psi(x,t)
\label{orig-Schr}
\end{align}
with a moving Dirichlet boundary condition:
\begin{align}
\label{dirch}
\psi(x=0,t)=\psi(x=L(t),t)=0.
\end{align}

 The expectation of the force acting on the wall is obtained by
\begin{equation}
\bar{F}=-\frac{\partial}{\partial L(t)} \langle \psi |H|\psi \rangle.
\label{averF}
\end{equation}
Noting $\frac{\partial}{\partial L} |\psi \rangle=\frac{1}{\dot{L}} \frac{\partial}{\partial t} |\psi \rangle=\frac{1}{i\hbar \dot{L}}H|\psi \rangle$ and its Hermitian conjugate, Eq. (\ref{averF}) reduces to
\begin{equation}
\bar{F}=- \langle \psi |\frac{\partial H}{\partial L(t)}|\psi \rangle.
\label{averF2}
\end{equation}
Hence the force operator is defined by
\begin{equation}
\hat{F}= -\frac{\partial H}{\partial L(t)}.
\label{opeF}
\end{equation}
However, the original Hamiltonian $H$ for the billiard with its time-dependent size $L(t)$  does not formally include
$L(t)$ explicitly. Therefore there is no way to define the force operator directly by using Eq.(\ref{opeF}).

To overcome this difficulty, we shall make the time-dependent canonical transformation of  $H$ related to the scale transformation of both the coordinate $x$ and
amplitude of the wave function $\psi$. This transformation, which was originally developed in the heat equation theory \cite{tai,mun}, is defined by \cite{raz83,raz91,cer99}
\begin{align}
H_1=e^{-iU}(H-i\hbar \frac{\partial}{\partial t} )e^{iU},
\end{align}
where
\begin{align}
U=-\frac{1}{2\hbar}(\hat{x}\hat{p}+\hat{p}\hat{x})\ln
L(t)=i\left(x\frac{\partial}{\partial x}+\frac{1}{2} \right)\ln L(t).
\end{align}
This canonical transformation  leads to the scaling of the coordinate $x$,
\begin{align}
e^{-iU} x e^{iU}=x L(t),
\end{align}
where on the right-hand side the new variable $x(\equiv y)$ varies in the range  $0 \le y \le 1$, which is time-independent!
Similarly the amplitude of the wave function is scaled as
\begin{align}
\tilde{\phi}(x,t)= e^{-iU}\psi(x,t)=\sqrt{L}\psi(xL,t),
\end{align}
so that the normalization factor of $\tilde{\phi}(x,t)$ becomes $L$-independent.
Finally the Schr\"odinger equation is transformed to
\begin{align}
i\hbar \frac{\partial \tilde{\phi}}{\partial t}=H_1\tilde{\phi}
\label{scaled_Schr}
\end{align}
with the new Hamiltonian
\begin{align}
H_1= -\frac{1}{2L^2}\frac{\partial^2 }{\partial x^2}+
i\hbar\frac{\dot{L}}{L}x\frac{\partial }{\partial x}+\frac{i\hbar}{2} \frac{\dot{L}}{L}.
\label{scaled_H1}
\end{align}
$\tilde{\phi}(x,t)$  now satisfies the fixed Dirichlet boundary condition $\tilde{\phi}(0,t)=\tilde{\phi}(1,t)=0$.
Equation (\ref{scaled_Schr}) with (\ref{scaled_H1}) is also available simply by replacing $\psi$ and $x$ and by$\frac{1}{\sqrt{L}}\tilde{\phi} $ and $xL$ respectively  in Eq.(\ref{orig-Schr}) with Eq.(\ref{orig-H}).

Taking $L$ derivative of $H_1$, we can rigorously define the force operator in the transformed space as
\begin{align}
\label{pseudo-f}
\tilde{F}&=-\frac{\partial H_1}{\partial L}=-\frac{1}{L^3}\frac{\partial^2}{\partial
x^2}+\frac{i\hbar\dot{L}}{L^2}\left(x\frac{\partial}{\partial x}+\frac{1}{2} \right)\nonumber \\
&\equiv\frac{1}{L^3}\tilde{p}_x^2-\frac{\dot{L}}{2L^2}\left(\tilde{x}\tilde{p}_x+\tilde{p}_x\tilde{x} \right).
\end{align}

Now, carrying out the inverse canonical transformation ($xL \to x$, etc.),
we have the force operator expressed in the original space as
\begin{align}
\label{exact-f}
\hat{F}=&e^{iU} \tilde{F}e^{-iU}=-\frac{1}{L}\frac{\partial^2}{\partial x^2}+
\frac{i\hbar\dot{L}}{L^2}\left(x\frac{\partial}{\partial x}+\frac{1}{2} \right)\nonumber\\
&=\frac{\hat{p}^2}{L} - \frac{\dot{L}}{2L^2} (\hat{x}\hat{p} +\hat{p}\hat{x}),
\end{align}
which certainly satisfies:
\begin{align}
\langle \psi |\hat{F}|\psi \rangle= \langle \tilde{\phi} |\tilde{F}|\tilde{\phi} \rangle.
\end{align}
The non-adiabatic term in Eq.~(\ref{exact-f}) agrees exactly with the phenomenological result
in Eq.~(\ref{phenom}).

\section{Perturbative theory of nonadiabatic force} \label{perturb}

In this Section we shall investigate the expectation of the force operator in Section \ref{f-operator} perturbatively with use of von Neumann equation for the density operator and adiabatic bases. The method is an extension of the Greenwood's linear response theory  \cite{green}.
Let' s assume that the cavity wall is fixed with the cavity size $L_0$ until the time origin $t=0$ and that  it suddenly begins to move with constant velocity $\dot{L}$ at $t>0$.

In the equilibrium statistical mechanics, the expectation of a given observable $\hat{O}$ is defined in energy-diagonal representation, as
\begin{align}
\langle \hat{O} \rangle={\rm Tr}\left(\hat{O} \frac{1}{e^{\beta(H-\mu)+1}} \right)=\sum_n O_{nn}\frac{1}{e^{\beta(E_n-\mu)+1}}.
\end{align}

In the near-equilibrium, the expectation value is evaluated in terms of the density operator $\rho$
\begin{align}
\rho=\sum_\alpha |\alpha\rangle \omega_\alpha \langle \alpha|,
\end{align}
with $\sum_\alpha \omega_\alpha=1$, as
\begin{align}
\langle \hat{O} \rangle= {\rm Tr}(\rho \hat{O})=\sum_\alpha \omega_\alpha O_{\alpha \alpha}.
\end{align}

We shall employ the original Hamiltonian $H$ and coordinate $x$. The density operator
$\rho$ for the Fermi gas obeys von Neumann equation
\begin{align}
i\hbar \frac{\partial \rho}{\partial  t}=\left[H,\rho\right].
\end{align}
With use of adiabatic basis ${|n \rangle }$, the matrix elements of $\rho$
satisfies
\begin{align}
\dot{\rho}_{nm}=\langle n | \frac{\partial \rho}{\partial  t}| m\rangle
+\langle \dot{n} | \rho | m\rangle+\langle n | \rho | \dot{m}\rangle,
\end{align}
with
\begin{align}
\langle \dot{n} | \rho | m\rangle+\langle n | \rho | \dot{m}\rangle=\sum_{\ell}
\left(\langle \dot{n} | l\rangle \rho_{\ell m}+ \rho_{n \ell} \langle \ell | \dot{m} \rangle
\right).
\end{align}

For the system under consideration, the instantaneous (adiabatic) eigenvalue problem is
given by
\begin{align}
\label{inst_eig_pr}
H(t)\psi_n\left(\equiv -\frac{1}{2}\frac{\partial^2 }{\partial x^2}\psi_n(t)
\right)=E_n \psi_n(t),
\end{align}
with adiabatic eigenstates and eigenvalues
\begin{subequations}
\label{adia-base}
\begin{align}
\psi_n\left(\equiv \langle x|n \rangle
\right)=\sqrt{\frac{2}{L(t)}}\sin\left(\frac{n\pi x}{L(t)} \right), \\
E_n=\frac{n^2\pi^2}{2L^2(t)},
\end{align}
\end{subequations}
where we prescribed $\hbar^2/m=1$.

Using Eq.~(\ref{inst_eig_pr}), we can obtain the following formulas
\begin{subequations}
\begin{align}
\langle \ell | \dot{m} \rangle=-\frac{\dot{L}}{L}\gamma_{\ell m}, \\
\langle \dot{n}|\ell \rangle=\langle \ell | \dot{n} \rangle^*=-\frac{\dot{L}}{L}\gamma_{\ell n},
\end{align}
\end{subequations}
where
\begin{align}
\label{matrix_el} \gamma_{\ell m}\equiv (-1)^{\ell+m+1}\frac{2\ell
m}{\ell^2-m^2}\left(1-\delta_{\ell m} \right).
\end{align}
Noting the pure-real nature of the adiabatic states in Eq.(\ref{adia-base}),
we see  $\langle \ell| \dot{\ell} \rangle=\langle \dot{\ell} | \ell \rangle=0$.
Hence we can put the diagonal element
$\gamma_{\ell \ell}=0$  in Eq.~(\ref{matrix_el}). The von Neumann equation now becomes
\begin{align}
\label{von_neumann2}
\dot{\rho}_{nm}=\frac{1}{i \hbar}\left(E_n-E_m\right)\rho_{nm}-\frac{\dot{L}}{L}\left(
\sum_{l\neq n}\gamma_{\ell n}\rho_{\ell m}+
\sum_{l\neq m}\gamma_{\ell m}\rho_{n\ell}
\right).
\end{align}

Eq.~(\ref{von_neumann2}) can be solved perturbatively. Let' s assume the solution to be expanded in $O(\dot{L}/L)$ as,
\begin{align}
\rho=f(H)+g_1\frac{\dot{L}(0)}{L(0)}+g_2\left(\frac{\dot{L}(0)}{L(0)}\right)^2+\dots.
\end{align}
with $f(H)=\frac{1}{e^{\beta(H-\mu)}+1}$.

One sees, for $O(1)$,
\begin{align}
\dot{f}_{nm}=0 \qquad (n=m).
\end{align}
Therefore
\begin{align}
\label{fermi_dis1}
f_{nm}=\frac{1}{e^{\beta(E_n(0)-\mu)}+1}\delta_{nm}\equiv f_n \delta_{nm}.
\end{align}
Then, for $O(\dot{L}/L)$, one sees
\begin{align}
\dot{g}_{1nm}=\frac{E_n-E_m}{i\hbar}g_{1nm}-(\gamma_{mn}f_m+\gamma_{nm}f_n),
\label{difeq_g1}
\end{align}
where the result in Eq.~(\ref{fermi_dis1}) was used. The solution of
Eq.~(\ref{difeq_g1}) is given by
\begin{align}
\label{solution_g1}
g_{1nm}=\frac{i\hbar \gamma_{mn}}{E_n-E_m} \left(1-e^{\frac{E_n-E_m}{i\hbar}t}
\right)(f_n-f_m).
\end{align}
For a correction of $O((\dot{L}/L)^2)$, the dominant contribution comes from the diagonal term satisfying
\begin{align}
\dot{g}_{2nn}= -\sum_\ell \gamma_{\ell n}(g_{1\ell n}+g_{1n\ell}).
\label{scor_nn}
\end{align}
With use of Eq. (\ref{solution_g1}),
\begin{align}
g_{1\ell n}+g_{1n\ell}=-2\gamma_{\ell n}
(f_n-f_l)\frac{\hbar}{E_n-E_l}\sin\frac{E_n-E_l}{\hbar}t .
\label{e2}
\end{align}
So, using Eqs.~(\ref{scor_nn}) and (\ref{e2}), we obtain
\begin{align}
\label{solution_g2} g_{2nn}=-2\sum_{l}
\gamma_{nl}^2&(f_n-f_l)\left(\frac{\hbar}{E_n-E_l}\right)^2\times\nonumber \\
&\left(1-\cos\frac{E_n-E_l}{\hbar}t \right).
\end{align}

Now let' s calculate the matrix elements of the force operator in Eq.~(\ref{exact-f}).
Using the adiabatic bases, we find
\begin{align}
\label{mat_el_force}
F_{mn}&=\frac{2}{L(t)}\int_0^{L(t)}\sin\left(\frac{m\pi x}{L(t)}
\right)\hat{F}\sin\left(\frac{n\pi x}{L(t)} \right)dx \nonumber \\
&=
\frac{(n\pi)^2}{L^3(t)}\delta_{mn}+\frac{i\dot{L}(t)}{L^2(t)}\gamma_{mn}.
\end{align}

Combining Eq.~(\ref{mat_el_force}) with Eqs.~(\ref{fermi_dis1}), (\ref{solution_g1}) and (\ref{solution_g2}),
the expectation value of the force operator becomes
\begin{align}
\label{average_force}
\bar{F}&=\langle \hat{F} \rangle={\rm Tr} (\rho \hat{F})=\sum_{m,n}\rho_{nm}F_{mn}=S_1+S_2+S_3,
\end{align}

where
\begin{align}
S_1=\sum_{n}f_{n}F_{nn}=\sum_n  \frac{(n\pi)^2}{L^3(t)}f_n,\qquad \qquad \qquad
\label{force-s1}
\end{align}
\begin{align}
S_2&=(\dot{L}/L)\sum_{m\neq n}g_{1nm}F_{mn}=\hbar\frac{\dot{L}(t)\dot{L}(0)}{L^2(t)L(0)}\times \nonumber\\
&\sum_{m\neq n}\gamma_{nm}\gamma_{mn}
\frac{f_n-f_m}{E_n-E_m}\left(1-e^{\frac{E_n-E_m}{i\hbar}t} \right),
\label{force-s2}
\end{align}
\begin{align}
S_3&=(\dot{L}/L)^2\sum_{n}g_{2nn}F_{nn}=-2\frac{\pi^2(\dot L(0))^2}{(L(0))^2L^3}\times \nonumber\\
&\sum_n n^2\sum_{l}
\gamma_{nl}^2(f_n-f_l)\left(1-\cos\frac{E_n-E_l}{\hbar}t
\right)\times\nonumber\\
&\left(\frac{\hbar}{E_n-E_l}\right)^2. \label{force-s3}
\end{align}

$S_1$ in Eq.~(\ref{force-s1}) gives rise to the expression for the
adiabatic force at finite temperature
\begin{align}
\label{average_force1}
\bar{F}_{ad}=
\sum_n  \frac{(n\pi)^2}{L^3(0)}
\frac{1}{e^{\beta(E_n(0)-\mu)}+1},
\end{align}
while $S_2$ in Eq.~(\ref{force-s2}) and $S_3$ in Eq.~(\ref{force-s3}) contribute to the nonadiabatic force.
In Eq.~(\ref{average_force1}), we have assumed the time range lies in
\begin{align}
\frac{\hbar}{\Delta E} \ll t \ll \frac{L}{\dot{L}},
\label{time-range}
\end{align}
where
the lower and upper limits of the inequality in Eq.~(\ref{time-range}) imply the minimum resolution of time
and the time necessary for the wall to move by order of $L$, respectively. This is a physically
imposed assumption, which will also be employed below.

$S_2$ can be rewritten, using
$\gamma_{nm}\gamma_{mn}=-\frac{(4mn)^2}{(m^2-n^2)^2}$ and
$L^2(t)(E_n-E_m)=\frac{\pi^2}{2}(n^2-m^2)$ and noting the fact
that, for a symmetric function $s(n,m)$, $\sum_{m\neq
n}s(n,m)\left(1-e^{\frac{E_n-E_m}{i\hbar}t} \right) =2\sum_{m\neq
n}s(n,m)\sin^2\left(\frac{E_n-E_m}{2\hbar}t\right)\to\sum_{m\neq
n}s(n,m)$, where  the final reduction is possible under the
assumption in Eq.~(\ref{time-range}). As a result, $S_2$ becomes
\begin{align}
\label{nonad_force1}
S_2=\frac{16\hbar^2\dot{L}(0)^2}{\pi^2L(0)}\sum_{n >m}\left[
\frac{m^2n^2}{(n^2-m^2)^3} (f_n-f_m)\right].
\end{align}
The factor $(f(E_n)-f(E_m))$ in Eq.~(\ref{nonad_force1}) gives a constraint under which
the summation over $m$ and $n$ should be taken.
Similarly $S_3$ becomes
\begin{align}
\label{nonad_force2} S_3=\frac{32\hbar^2(\dot
L(0))^2}{\pi^2(L(0))}\sum_{n >m}
\left[\frac{m^2n^2}{(n^2-m^2)^3} (f_n-f_m)\right].
\end{align}
To summarize, the nonadiabatic force is given by
\begin{align}
\label{nonad_force}
\bar{F}_{non-ad}=S_2+S_3=C\frac{(\dot L(0))^2}{(L(0))}
\end{align}
with
\begin{align}
\label{coeff-C}
C=\frac{48\hbar^2}{\pi^2}\sum_{n > m}\left[
\frac{m^2n^2}{(n^2-m^2)^3}
(f_n-f_m)\right],
\end{align}
where $C<0$ since Fermi distribution function monotonically decreases with energy and $ f_n-f_m \equiv f(E_n)-f(E_m) <0$ for $n>m$.
$\bar{F}_{non-ad}$ in Eq.~(\ref{nonad_force}) is proportional to $\dot{L}^2$ and does not break the time-reversal symmetry,
in marked contrast with the general
conjecture so far.  The result in Eqs.~(\ref{average_force1}) and (\ref{nonad_force}) will be confirmed by the exact analytical result
in the next Section.

Here we should give two comments:

i) The first one is concerned with the level crossing. From an
experimental viewpoint, we are considering a quasi-one-dimensional
(1-d) hard- or soft-walled rectangular parallelepiped. In this
case, the energy gaps between sub-bands are large enough  not to
meet crossings among sub-bands. Therefore the dynamics within each
sub-bands (e.g., the lowest sub-band) used in our scheme is
guaranteed. As a more general case, one might consider a 3-d
rectangular parallelepiped with the size $L_x \times L_y \times
L_z$, one of whose walls is moving in $x$-direction. Then each
adiabatic state is characterized by a set of quantum numbers
$(n_x,n_y,n_z)$ and the energy spectra as a function of $L_x$
might show level crossings among manifolds with different
$n_x,n_y$ and $n_z$. If a confined particle is initially in a
manifold with the fixed $n_y$ and $n_z$ and the cavity expands
only in $x$-direction, however, there occurs no transition among
manifolds with different $n_y$ and $n_z$ and thereby energy
crossings do not affect the present dynamics at all. Finally, one
can conceive an expanding 3-d spherical billiard, which has level
crossings among manifolds with different angular momenta. Since
there is no transition matrix element among different angular
momentum states in the symmetry-keeping dynamics, however, the
dynamics is free from the problem of level crossings if a
zero-angular momentum state will be chosen as an initial state,
which again guarantees our scheme.

ii) The second one is whether or not the expression for the force
operator and the expectation of the non-adiabatic force quadratic
in the rate of dilation under the "time-dependent" Dirichlet
boundary condition (TDD)  would be available directly from a
variational method. Berry and Wilkinson (BW) \cite{BW} considered
the static (adiabatic) eigenvalue problem of a  triangular
billiard under the "time-independent" Dirichlet boundary condition
(TID), to study the density of diabolical points and both shifts
and splitting of level degeneracies. They had recourse to a
degenerate perturbation theory with use of diabatic eigenstates at
the degenerating point, indicating (in Appendix of their paper)
that the off-diagonal energy matrix elements  are zero for
dilations. It is not easy to interpret the present dynamical
result under TDD  only in terms of the static one of BW under TID.
The present work is concerned with a one-dimensional billiard with
a time-dependent walls, where the adiabatic eigenvalues have no
degeneracy and only the level shifts occur against the adiabatic
dilation. In the context of the adiabatic (AF) and non-adiabatic
forces (NAF) which are given  on the second line in
Eq.(\ref{mat_el_force}), the non-diagonal matrix elements of AF
certainly vanish, consistent with BW, but matrix elements of NAF
are new, whose counterpart cannot be found in the treatment of BW
under TID. The force operator of BW is only concerned with AF.
Nonvanishing matrix  elements of  NAF  in Eq.(\ref{mat_el_force})
are due to a dynamical contribution in Eq.(\ref{exact-f}) coming
from TDD. A mechanism of the absence of a term linear in the rate
of dilation in the expectation value of the force operator in
Eq.(\ref{average_force}) is not due to vanishing non-diagonal
matrix elements of AF and can not be explained directly within the
static framework under TID. It is caused by a subtle cancellation
of  the linear cross-coupling terms among the matrix elements of
the force operator expressed as a series expansion w.r.t  the rate
of dilation and the density matrix expressed in the similar
expansions in the framework of the extended Kubo-Greenwood
formula. On the other hand, Berry and Klein \cite{BK} were once
involved in the similar subject as the present one, but they
showed neither the definition of NAF operator nor the expectation
value of the force as a power series in the rate of change of the
scale size of the container.

\section{Exact analysis} \label{exact}

Exact solution of the Schr\"odinger equation with a moving Dirichlet boundary condition
due to the motion of a wall was found by Makowski et.al \cite{mak91,mak92}. The greatness of their work lies in that
they discovered the transitionless basis functions where the adiabatic states are also the solution of the time-dependent
Schr\"odinger equation,
which recently received a great attention in the context of the shortcut to the adiabatic dynamics \cite{chen}.
With use of their basis functions we can proceed to
evaluate the nonadiabatic force exactly.
After a brief summary of their results, we shall carry out this procedure.

The system we are going to explore is described by the Schr\"odinger equation
\begin{align}
i\hbar \frac{\partial \psi(x,t)}{\partial t}=
-\frac{1}{2}\frac{\partial^2\psi(x,t)}{\partial x^2}
\label{schrad}
\end{align}
where the wave function satisfies the moving Dirichlet boundary condition in Eq. (\ref{dirch}).

After the scaling of space coordinate $x$ and wave function $\psi$ by $L(t)$ and
$\sqrt{L(t)}$, respectively as
\begin{align}
x\to y=\frac{x}{L(t)} \nonumber\\
\psi\to\tilde{\phi}=\sqrt{L(t)}\psi,
\label{scalings}
\end{align}
Schr\"odinger equation with the moving boundary becomes
\begin{align}
i\hbar \frac{\partial \tilde{\phi}}{\partial t}=-\frac{1}{2L^2}\frac{\partial^2 \tilde{\phi}}{\partial y^2}+
i\hbar\frac{\dot{L}}{L}y\frac{\partial \tilde{\phi}}{\partial y}+\frac{i\hbar}{2} \frac{\dot{L}}{L}\tilde{\phi}
\label{scaled_eq1}
\end{align}
with a static Dirichlet boundary condition, i.e.,
\begin{align}
\tilde{\phi}(0)=\tilde{\phi}(1)=0.\nonumber
\end{align}
The transformation above is nothing but the time-dependent canonical transformation described in Section \ref{f-operator}.

Then, applying the gauge transformation
\begin{align}
\phi(y,t)=G\tilde{\phi}(y,t)=\exp\left(-\frac{i\hbar}{2} \dot{L}(t)L(t)y^2\right)\tilde{\phi}(y,t),
\label{gauge_trans}
\end{align}
Eq.~\re{scaled_eq1} can be reduced to the Schr\"odinger equation for the time-dependent harmonic oscillator:
\begin{align}
i\hbar L^2\frac{\partial \phi}{\partial t}=-\frac{1}{2}\frac{\partial^2 \phi}{\partial y^2} +\frac{\hbar^2}{2}L^3\ddot Ly^2\phi .
\label{eq18}
\end{align}

Eq.~(\ref{eq18}) can be solved exactly if
the time-dependence of the boundary satisfies the following equation \cite{mak91}:
\begin{align}
L^3\ddot L=const =-\frac{1}{4}B^2.
\label{A}
\end{align}

For a linearly expanding or contracting billiard with the constant wall velocity $\dot{L}=\dot{L}(0)$, i.e.,
\begin{align}
\label{v-length}
L(t)=L_0+\dot{L}t,
\end{align}
the condition (\ref{A}) is satisfied, $B=0$. (A general case of $B \ne 0$ will be investigated in Appendix \ref{nonconst_vel}.)

The solution in this case is
\begin{align}
\psi_n(x,t)&=\sqrt{\frac{2}{L(t)}}\exp\left(\frac{i\hbar\dot{L}}{2L}x^2-\frac{in^2\pi^2}{2\hbar}
\tau(t)\right) \nonumber \\
&\times\sin\left(\frac{n\pi x}{L(t)} \right),
\label{exact_solution}
\end{align}
where $\tau(t)$ is a new time variable defined by
\begin{align}
\label{scaled-time}
\tau(t)\equiv\int_0^t\frac{ds}{L^2(s)}.
\end{align}
The solution
(\ref{exact_solution}) is the transitionless state where the adiabatic state also serves as the solution
of the time-dependent Schr\"odinger equation,
which recently received a renewed attention \cite{chen}.
An assembly of states in Eq.~(\ref{exact_solution}) constitute the complete ortho-normal set.

Let's obtain the adiabatic and nonadibatic forces acting on the moving wall which is confining the Fermi gas
into the cavity, by evaluating the expectation of Hamiltonian. Statistical weight factors (Fermi distribution)
will be incorporated a posteriori.
In the case of a linearly expanding cavity described by Eq.~(\ref{v-length}), the initial state of a particle is given
by $\sqrt{\frac{2}{L_0}}\sin\left(\frac{l\pi
x}{L_0}\right)$ with the  eigenvalue $E_l(0)=\frac{l^2\pi^2}{2L_0^2}$ and the wall suddenly begins
to move with constant velocity $\dot{L}$.
The solution of Eq.~\re{schrad} can be expressed in terms of the transitionless states in
Eq.~\re{exact_solution} as
\begin{align}
\psi=&\frac{1}{\sqrt{L}}\exp\left(\frac{i\hbar}{2}\dot{L}Ly^2\right)\phi,\nonumber \\
\phi&=\displaystyle\sum_{n}{c_n(t)\varphi_n}(y), \qquad \varphi_n(y)=\sqrt{2}\sin(n\pi
y),
\end{align}
where expansion coefficients are given by
\begin{align}
c^{(\ell)}_{n}(t)=&c^{(\ell)}_{n}(0)\exp\left(-\frac{in^2\pi^2}{2\hbar} \tau(t)
\right)
\end{align}
with
\begin{align}
c^{(\ell)}_n(0)=2\int_{0}^{1}\sin{(l\pi y)}\sin{(n\pi
y)}\exp{\left(-\frac{i\hbar}{2}\dot{L}(0)L(0)y^2 \right)}dy.
\label{initial_coeffs}
\end{align}

 The average energy can be represented as
\begin{align}
\langle E^{(\ell)}(t)\rangle&=\int_{0}^{L(t)}\psi^*(x,t)H\psi(x,t)dx\nonumber \\
&=\frac{1}{L}\int_{0}^{1}\psi^*(y,t)
\left(-\frac{\partial^2}{2\partial y^2} \right)\psi(y,t)dy \nonumber\\
&=\frac{1}{2L^2}I_0+
\frac{\hbar\dot{L}}{L}{\rm Im}(I_1)+\frac{\hbar^2 \dot{L}^2}{2}I_2,
\label{energy_main}
\end{align}
where $I_0,I_1$ and $I_2$ are respectively defined by
\begin{align}
I_0=&\int_{0}^{1}\mid\phi_y\mid^2 dy \nonumber \\
=&2\pi^2
\sum_{n^{\prime}}c^{(\ell)*}_{n^{\prime}}(t)n^{\prime}\sum_{n}c^{(\ell)}_{n}(t)n\int_{0}^{1}\cos(n^{\prime}\pi y)\cos(n\pi y)dy \nonumber \\
=&\pi^2
\sum_{n}\mid c^{(\ell)}_n(0)\mid ^2 n^2,
\label{i00}
\end{align}
\begin{align}
I_1(t)=\int_{0}^{1}y\phi^*\phi_ydy=
\sum_{n'}c^{(\ell)*}_{n'}(t)\sum_{n}c^{(\ell)}_{n}(t)n\pi J_1(n,n'),
\label{i01}
\end{align}
\begin{align}
I_2(t)=\int_{0}^{1}y^2\mid\phi\mid^2 dy=
\sum_{n'}c^{(\ell)*}_{n'}(t)\sum_{n}c^{(\ell)}_{n}(t) J_2(n,n'),
\label{i02}
\end{align}
with
$J_1(n,n')$ and $J_2(n,n')$ given in Appendix \ref{IJ}.

Since the work done by the force $F^{(\ell)}(t)$ (on the moving wall) from a Fermi particle is supplied
by the excess energy induced by its energy loss, we find
\begin{align}
\bar{F}^{(\ell)}(t)=-\frac{\partial \langle E^{(\ell)}(t)\rangle}{\partial
L}=\frac{1}{L^3}I_0 +\frac{\hbar \dot{L}}{L^2}{\rm Im}(I_1).
\label{force-11}
\end{align}

The force can also be calculated by taking the expectation of the force operator $\hat{F}$ in Eq.~(\ref{exact-f})
defined in the original space:
\begin{align}
\bar{F}(t)=&\int_{0}^{L(t)}\psi^*(x,t)\hat{F}\psi(x,t)dx
=\frac{1}{L^3}I_0 +\frac{\hbar\dot{L}}{L^2}{\rm
Im}(I_1) \nonumber \\
&\equiv\bar{F}_{ad}+\bar{F}_{non-ad}.
\label{force-12}
\end{align}
Eqs.~(\ref{force-11}) and (\ref{force-12}) exactly agree mutually, which guarantees the validity
of the force operator defined in Section \ref{f-operator}.

To investigate the nature of the nonadiabatic force more carefully, however, we
must estimate the integrals $I_0$ and $I_1$. By expanding the exponential in Eq.~(\ref{initial_coeffs}) as
\begin{align}
\exp{\left(-\frac{i\hbar}{2}\dot{L}(0)L(0)y^2
\right)}=&1-\frac{i\hbar}{2}\dot{L}(0)L(0)y^2\nonumber \\
&-\frac{\hbar^2}{8}(\dot{L}(0)L(0))^2y^4+\dots,
\end{align}
we find (see Appendix \ref{IJ})
\begin{align}
\label{i0-exp}
I_0=\pi^2\ell^2+ \frac{\pi^2 \hbar^2(\dot{L}(0)L(0))^2}{4}& \left(\sum_n n^2(J_2(n,\ell))^2 - \ell^2J_3(\ell,\ell)\right)
\end{align}
and
\begin{align}
\label{i1-exp}
{\rm Im}(I_1)=&-\frac{\pi\hbar}{2}\dot{L}(0)L(0)\sum_n 2nJ_1(n,\ell) J_2(n,\ell) \nonumber \\
&=-\hbar\dot{L}(0)L(0)\sum_{n\neq \ell}\left(\frac{16}{\pi^2} \right)
\frac{(n\ell)^2}{(n^2-\ell^2)^3}\nonumber \\
& \times \cos\left(\frac{\pi^2}{2}(\ell^2-n^2)\tau(t) \right).
\end{align}
In Eq.(\ref{i1-exp}) the last factor can be taken as $ \cos(\cdots)\sim 1$ in the time range in Eq.(\ref{time-range}).
We find that the $\dot{L}(0)$-dependent terms are included not only in $I_1$ but also in $I_0$.
Substituting Eqs.~(\ref{i0-exp}) and (\ref{i1-exp}) into Eq.~(\ref{force-12}), we see:
(i) the $\dot{L}(0)$-independent term in  Eq.~(\ref{force-12}) gives rise to the adiabatic force $(\bar{F}_{ad})$;
(ii) the remaining terms give the nonadiabatic force $(\bar{F}_{non-ad})$.

Picking up the first term on r.h.s of Eq.~(\ref{i0-exp}), multiplying statistical weight $f_n$ and summing up over all initial eigenstates,
we find
\begin{align}
\label{adiab_force1}
\bar{F}_{ad}=
\sum_n  \frac{n^2\pi^2}{L^3(0)}f_n,
\end{align}
which justifies Eq.~(\ref{average_force1}).

Taking together Eq.~(\ref{i1-exp}) and the second term on r.h.s. of Eq.~(\ref{i0-exp}), multiplying statistical weight $f_n$ and summing up over all
initial eigenstates, we find
\begin{align}
\label{ex-nonad1}
\bar{F}_{non-ad}=C^{\prime}\frac{(\dot{L}(0))^2}{L(0)}
\end{align}
with
\begin{align}
\label{coeff-2}
C^{\prime}=
&\sum_{n}\left[\sum_{m(\ne n)}\frac{16\hbar^2}{\pi^2}\left(\frac{m^2 n^2}{(m^2-n^2)^3}+\frac
{m^4n^2}{(m^2-n^2)^4}\right)\right.\nonumber\\
&\left.-\frac{\hbar^2\pi^2}{4}\left(\frac{n^2}{5}-\frac{1}{\pi^2}+\frac{3}{2n^2\pi^4}\right)\right]f_n,
\end{align}
which is again negative due to the dominant term proportional to $n^2$.  Irrespective of the direction of the moving wall, the non-adiabatic force always acts inwards and is proportional to the square of the wall velocity, which is in marked contrast with
the general belief that the non-adiabatic force should be linear in the wall velocity and mimic the irreversible or frictional force.
There is a minor discrepancy between the absolute values of $C$ and $C^{\prime}$, which is due to the difference in the way of solving the problem \cite{comm}.

In closing this Section, we should note the following two remarks:

i) Firstly there is a mystery in obtaining the nonadiabatic force in the exact analysis above.
In this Section we had recourse to the transitionless states as basis functions.
We can see neither nonadiabatic transition nor nonadiabatic force so long as tracking individual transitionless states.
In fact, if we shall evaluate the expectation value $\bar{F}=\langle \hat{F} \rangle$ using only a single transitonless state in
Eq.~(\ref{exact_solution}), we will obtain  formally the same result as in Eq.~(\ref{force-12})
but with $I_0=\pi^2\ell^2$ and ${\rm Im}(I_1)=0$ and can see no nonadiabatic force.
Throughout this paper, we are considering the case that the wall is fixed up to the initial time $t=0$
and suddenly moves at $t>0$. Therefore the eigenstate under the fixed boundary generates at $t=0$ a mixture
of the transitionless states that are eigenstates of the moving boundary, giving rise to nonvanishing coefficients $\{c_n\}$.
Exploring  Eqs.~(\ref{i00}) and (\ref{i01}), we can understand that the correlation among non-zero
coefficients $\{c_n\}$ resulted in
the nonvanishing nonadiabatic force in Eq.~(\ref{ex-nonad1}). By contrast, it is quite easy to see the mechanism for
nonadiabatic force in the perturbative theory of Section \ref{perturb}, where the energy diffusion among
standard adiabatic states can explain the nonadiabatic force.

ii) Secondly the exact analysis can also reveal the nature of the non-adiabatic force in the case when the expanding rate of the cavity is not constant,
so long as $L(t)$ obeys Eq.(\ref{A}), namely when $L(t)=\sqrt{at^2+bt+c}$.  Under the initial condition with $L(0)=L_0$ and $\dot{L}(0)=\frac{b}{2\sqrt{c}}\ne 0$, $\bar{F}(t)$ can be calculated, leading to the identical result $F_{non-ad}=\bar C \frac{(\dot L(0))^2}{L(0)}$ with a negative constant $\bar{C}$. The details are given in Appendix \ref{nonconst_vel}. Thus our assertion that the non-adiabatic force is quadratic in the wall
velocity and thereby does not break the time-reversal symmetry does hold also for the hard-wall cavity with the time-dependent wall velocity.

\section{Case of soft-walled confinement} \label{soft}
\begin{figure}[htb]
\centerline{\includegraphics[width=\columnwidth]{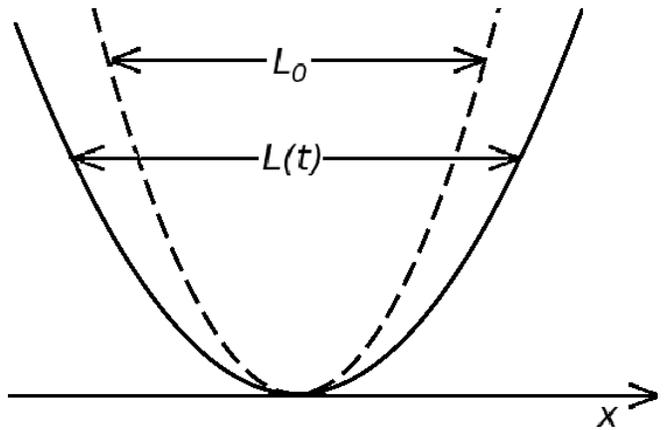}}
\caption{Soft wall confining the quantal gas. $L(t)$ is the time-dependent confining length.}
\label{soft-wall}
\end{figure}

To see the universality of our argument so far,
we proceed to investigate the case of the soft-wall confinement, and consider the force acting on the soft wall.
Here the Fermi gas is assumed to be confined in a harmonic trap with the confining length $L$ changing linearly
in time (see Fig.\ref{soft-wall}).

To evaluate the expectation of energy and force,
we shall solve the Schr\"odinger equation for a particle under the harmonic trap with the time-dependent trapping frequency,
\begin{align}
\label{hox}
i\hbar\frac{\partial \psi(x,t)}{\partial t}=H\psi(x,t)=-\frac{1}{2}\frac{\partial^2 \psi(x,t)}{\partial
x^2}+\frac{1}{2}m\omega^2(t)x^2\psi(x,t), \nonumber \\
-\infty<x<+\infty,
\end{align}
where, due to the prescription $\hbar^2/m=1$,  we see $m=\hbar^2$.
In Eq.(\ref{hox}) $\omega(t)$ is expressed in terms of the time-dependent confining length: $\hbar\omega(t)=\frac{1}{L^{2}(t)}$
where $L(t)=L_0+\dot{L}t$ with constant $\dot{L}$.

After scale and gauge transformations like Eqs. (\ref{scalings}) and (\ref{gauge_trans}), Eq.~(\ref{hox}) is reduced to
\begin{align}
\label{hoy3}
i\hbar L(t)^2\frac{\partial \phi(y,t)}{\partial t}=-\frac{1}{2}\frac{\partial^2 \phi(y,t)}{\partial y^2}+\frac{1}{2}\hbar^2 y^2\phi(y,t),
\end{align}
where $\psi(x,t)=\frac{1}{\sqrt {L(t)}}\exp(\frac{i}{2}\hbar\dot{L}(t)L(t)y^2)\phi(y,t)$ and $x=L(t) y$.
The solution of Eq.~(\ref{hoy3}) can be written as
\begin{align}
\label{psol}
\phi_n(y,t)=\exp\left(-i\int_0^t\frac{n+\frac{1}{2}}{\hbar L^2(s)}ds\right)Y_n(y),
\end{align}
where
\begin{align}
\label{hp}
Y_n(y)=\sqrt{\frac{\sqrt{\hbar}}{2^n
n!\sqrt{\pi}}}\exp\left(-\frac{\hbar y^2}{2}\right)H_n(\sqrt{\hbar}y)
\end{align}
with
\begin{align}
\label{hp2}
H_n(z)=(-1)^n e^{z^2}\frac{d^n}{dz^n}e^{-z^2}.
\end{align}
An assembly of solutions in Eq. (\ref{psol}), which are transitionless states, constitute the complete ortho-normal set.

As in the case of the hard-wall cavity, let's  require that the initial state is an eigenstate under the fixed harmonic trap:
\begin{align}
\label{initialx}
\psi(x,t)|_{t=0}=&\psi_0(x)=\frac{1}{\sqrt{L(0)}}Y_\ell\left(\frac{x}{L(0)} \right),
\end{align}
which is equivalent to
\begin{align}
\label{initialy}
\phi(y,0)=Y_\ell(y)\exp\left(-\frac{i}{2}\hbar\dot L(0)
L(0)y^2\right).
\end{align}
The time-dependent solution satisfying this initial condition is expressed as
\begin{align}
\label{gensol}
\phi(y,t)=\sum_{n=0}^{+\infty}c_n \phi_n(y,t), \end{align}
Expansion coefficients $c_n$ are determined by
\begin{align}
\label{gensol}
c_n&=\int_{-\infty}^{+\infty} \phi(y,0)Y^*_n(y)dy \nonumber \\
&= \int_{-\infty}^{+\infty}Y_\ell(y)Y_n(y)\exp(-\frac{i}{2}\hbar\dot
L(0)L(0)y^2)dy.
\end{align}

The average energy $\langle E(t)\rangle$ is calculated by
\begin{align}
\label{energy}
\langle E(t)\rangle&=\int_{-\infty}^{+\infty}\psi^*(x,t)H\psi(x,t)dx \nonumber \\
&=\frac{1}{2L^2}K_0+\frac{\hbar\dot L}{L}{\rm Im} (K_1)+\frac{\hbar^2\dot L^2}{2}K_2,
\end{align}
where
$K_0=\int_{-\infty}^{+\infty}\left(|\phi_y|^2+y^2|\phi|^2\right)$, $K_1=\int_{-\infty}^{+\infty}y\phi^*\phi_y dy$ and
$K_2=\int_{-\infty}^{+\infty}y^2|\phi|^2 dy$.

The average force is obtained by taking the derivative of $\langle E(t)\rangle$ with respect to $L$ as
\begin{align}
\label{soft-f1}
\bar{F}=-\frac{\partial E}{\partial L(t)}=
\frac{1}{L^3}K_0+\frac{\hbar\dot{L}}{L^2} {\rm Im}(K_1).
\end{align}

On the other hand,  one should evaluate the expectation value of the force operator to reproduce the above result.
Although the range of $x$ is not limited in the case of soft-wall confinement, one can define the force operator
using the time-dependent canonical transformations related to scaling of both coordinates and wave functions and its inverse transformations in Section \ref{f-operator}.
The force operators in the  original space is given by
\begin{align}
\label{soft-force-ori}
\hat{F}=-\frac{1}{L}\frac{\partial^2 }{\partial
x^2}+\frac{x^2}{L^3}+i\frac{\hbar \dot{L}}{L^2}\left(x\frac{\partial }{\partial x}+\frac{1}{2}\right),
\end{align}
which includes the second term missing in Eq.(\ref{exact-f}).
Therefore its expectation is
\begin{align}
\bar{F}=\langle \hat{F}\rangle=\frac{1}{L^3}K_0+\frac{\hbar \dot{L}}{L^2}{\rm Im}(K_1),
\end{align}
which accords with Eq.~(\ref{soft-f1}) and guarantees the validity the definition of the force operator in Eq.~(\ref{soft-force-ori}).

With use of asymptotic expressions for
$K_0$ and $K_1$,
we again find  $F_{non-ad}=C_{soft} \frac{\dot{L}(0)^2}{L(0)}$ with a coefficient $C_{soft} < 0$ , namely,
the nonadiabatic force is proportional to the square of  velocity of the soft wall, and never breaks the time-reversal symmetry.

\section{Conclusions} \label{concl}
We investigated the nature of the non-adiabatic force acting on the cavity wall, which is generated from the non-interacting  quantal gas confined in the expanding cavity.
Firstly, with use of the time-dependent canonical transformations by which we can move to the non-expanding cavity,  the force
operator is defined. Secondly, we analyzed the expectation of the force operator perturbatively with use of von Neumann equation for the density operator, which works when the cavity wall suddenly
begins to move at time origin. We found that the non-adiabatic force is quadratic in the wall
velocity and thereby does not break the time-reversal symmetry, in marked contrast with the existing conjecture.  Finally, using an assembly
of the transitionless quantum states, we obtain the  nonadiabatic force exactly. The exact result
justifies the validity of both the definition of force operator and the issue of the perturbative theory, and guarantees the present findings in the general case  when the expansion rate of the cavity is time-dependent.
The mysterious mechanism of nonadiabatic transition with use of transitionless quantum states
is also  explained. The study is done on both cases of  the hard-wall and soft-wall
confinement with the time-dependent confining length. Quantum fluctuation theorem, deviation from the standard Fermi-Dirac distribution and equation of states, etc. in the expanding cavity where Hilbert space is time-dependent also constitute interesting subjects, which will be investigated in due course.

{\em Acknowledgments.} We are grateful to S. Tanimura, A. Sugita and A. Terai  for useful comments.
K.N. expresses special thanks to B. Mehlig for enlightening discussions in the early stage of the present work and to M.V. Berry for kindly informing us of his old paper with Klein touching on the similar subject as the present one.  T.M. acknowledges a support under the JSPS program (Grant in Aid 22$\cdot$7744). The work is also supported through a project of the Uzbek Academy of Sciences (FA-F2-084).

\appendix
\label{append}

\section{The case $L(t)=\sqrt{at^2+bt+c}$}
 \label{nonconst_vel}
 We consider the expanding cavity of the size $L(t)$ governed by Eq.(\ref{A}), which has the general solution
 $L(t)=\sqrt{at^2+bt+c}$ with $B^2=b^2-4ac$ and the initial velocity $\dot{L}(0)=\frac{b}{2\sqrt{c}}\ne 0$.
 In this case the reduced Schr\"odinger
 equation in Eq.(\ref{eq18}) takes the following form:
\begin{align}
\label{ncv1}
i\hbar L^2\frac{\partial \phi}{\partial
t}=-\frac{1}{2}\frac{\partial^2 \phi}{\partial
y^2}-\frac{1}{8}\hbar^2B^2y^2\phi,
\end{align}
where $y=x/L(t)$ and
\begin{align}\label{ncv2}
\phi(y,t)=\sqrt{L}\exp\left(-\frac{i\hbar}{2}L\dot L
y^2\right)\psi(yL(t),t).
\end{align}

Changing the time variable from $t$ to $\tau$ defined by Eq.(\ref{scaled-time}), Eq.(\ref{ncv1}) can be solved as
\begin{align}
\label{ncv3}
\phi_n (y,t)=\varphi_n(y)\exp\left(\frac{i}{\hbar}K_n\tau(t) \right)
\end{align}
with
\begin{align}
\label{ncv4}\varphi_n(y)=A_n\exp\left(-\frac{By^2}{2}\right)y \
{\rm Re}\left[M\left(\frac{3}{4}+i\frac{K_n}{\hbar B},
\frac{3}{2}, \frac{i\hbar B}{2} y^2\right)\right],
\end{align}
where $M(a,b,y)$ is Kummer function (i.e., the confluent hypergeometric
function) and $K_n$ are the solutions of the equation
\begin{align}
\label{ncv5}
M\left(\frac{3}{4}+i\frac{K_n}{\hbar B}, \frac{3}{2}.
\frac{i\hbar B}{2}\right)=0.
\end{align}
In the semiclassical region where $\hbar B \sim 0$, one can see $K_n\sim\frac{n^2\pi^2}{2}$.

Now we shall solve the time-dependent problem under the initial state
\begin{align}
\label{ncv8}\phi(y,0)=\sqrt{2}\sin(\pi \ell y),
\end{align}
which corresponds to
$ \psi(x,0)=\sqrt\frac{2}{L(0)}\sin(\frac{\pi \ell x}{L(0)} )$
with the eigen-energy $E_{\ell}(0)=\frac {\ell^2\pi^2}{2L_0^2}$.
Expanding
$\phi(y,t)$ in terms of  the complete set of functions $\varphi_n(y)$, we
have
\begin{align}
\label{ncv9} \phi(y,t)=\sum_n c_n(t)\varphi_n(y),
\end{align}
where
\begin{align}
\label{ncv10} c_n(t)
=c_n(0)\exp\left(-\frac{i}{\hbar}K_n\tau(t)\right)
\end{align}
with $c_n(0)$  given by
\begin{align}
\label{ncv10}
c_n(0)&=\sqrt{2}\int\limits_0^1\varphi_n^*(y) \sin{\ell\pi
y}\exp\left(-\frac{i\hbar}{2}L(0)\dot L(0) y^2\right)dy \nonumber\\
= &\bar J_0(n,\ell)-\frac{i\hbar}{2}L(0)\dot L(0)\bar
J_2(n,\ell)\nonumber \\
-&\frac{\hbar^2}{8}(L(0)\dot L(0))^2\bar J_3(n,\ell)+ \cdots .
\end{align}
$\bar{J}_0$, $\bar{J}_2$ and $\bar{J}_3$ are defined respectively as
\begin{align}\label{barJ0}
\bar J_0(n,\ell)=\sqrt{2}\int\limits_0^1 \varphi_n^*(y)\sin(\pi
\ell y) dy,
\end{align}
\begin{align}\label{barJ2}
\bar J_2(n,\ell)=\sqrt{2}\int\limits_0^1y^2\varphi_n^*(y)\sin(\pi
\ell y) dy,
\end{align}
and
\begin{align}\label{}
\bar J_3(n,\ell)=\sqrt{2}\int\limits_0^1y^4\varphi_n^*(y)\sin(\pi
\ell y) dy.
\end{align}

Pair products of $c_n$ s are
\begin{align}
\label{mcc-0}
c_{n'}(0)& c_n^*(0)  \sim \bar J_0({n',\ell}) \bar
J_0({n,\ell})\nonumber\\
 -&\frac{i\hbar}{2}\dot{L}(0)L(0)\left[\bar J_0({n',\ell})
\bar J_2(n,\ell)-\bar J_0({n,\ell})\bar J_2(n',\ell)\right]  \nonumber\\
-&\frac{\hbar^2}{8}(\dot{L}(0)L(0))^2\left[\bar J_0({n',\ell})
\bar
J_3(n,\ell)\right. \nonumber\\
+&\bar J_0({n,\ell}) \bar J_3(n',\ell)+\left. 2\bar
J_2(n',\ell)\bar J_2(n,\ell) \right]
\end{align}
and
\begin{align}\label{ct}
c_{n'}(t) c_n^*(t)= c_{n'}(0)
c_n^*(0)\exp\left(-i\int\limits_0^t\frac{K_{n^{\prime}}-K_n}{\hbar
L^2(s)}ds\right).
\end{align}

The expectation for the force operator in Eq.(\ref{exact-f})  is obtained as
\begin{align}
\langle F^{(\ell)}(t)\rangle&=\int_{0}^{L(t)}\psi^*(x,t)\hat{F}\psi(x,t)dx\nonumber \\
&=\frac{1}{L^3}\bar I_0+ \frac{\hbar\dot{L}}{L^2}{\rm Im}(\bar
I_1),
\label{energy_main}
\end{align}
with $\bar{I}_0$ and $\bar{I}_1$ being expressed as
\begin{align}\label{I0bar}
&\bar I_0= \int\limits_0^1|\phi_y|^2dy=2\sum_n |c_n(0)|^2K_n\nonumber\\
&\qquad +\frac{1}{4}\hbar^2B^2\sum_n\sum_{n'}{\rm
Re}\left[c_n^*(t)c_{n'}(t)\right]\int_0^1\varphi_n\varphi_{n'}y^2dy
\end{align}
and
\begin{align}\label{I1bar}
\bar I_1&=\int\limits_0^1 y\phi^*\phi_y dy\nonumber\\
&=\sum_n\sum_{n'}
c_n^*(t)c_{n'}(t)\int_0^1y\varphi_n^*\frac{\partial\varphi_n}{\partial y} dy,
\end{align}
respectively.
Using the asymptotic forms for the pair products in Eq.(\ref{mcc-0}) together with Eq.(\ref{ct}) and taking the short-time region employed in the main text, we reach the result:
\begin{align}\label{asymp_force}
F_{non-ad}=\bar C \frac{(\dot L(0))^2}{L(0)},
\end{align}
which is again proportional to the square of the wall velocity.

\section{Calculation of $I$ and $J$}
\label{IJ}
Coefficients $c^{(\ell)}_n(0)$ are defined by
\begin{align}
c^{(\ell)}_n(0)& \nonumber\\
=&2\int_{0}^{1}\sin{(l\pi y)}\sin{(n\pi
y)}\exp{\left(-\frac{i\hbar}{2}\dot{L}(0)L(0)y^2
\right)}dy\nonumber\\
=&\delta_{n\ell}-\frac{i\hbar}{2}\dot{L}(0)L(0)J_2(n,\ell)\nonumber\\
&-\frac{\hbar^2}{8}(\dot{L}(0)L(0))^2J_3(n,\ell)+\cdots.
\end{align}
Their products are
\begin{align}
c^{(\ell)*}_{n'}(0) & c^{(\ell)}_n(0)  \sim \delta_{n'\ell} \delta_{n\ell}\nonumber\\
 &-\frac{i\hbar}{2}\dot{L}(0)L(0)\left[\delta_{n'\ell}
J_2(n,\ell)-\delta_{n\ell} J_2(n',\ell)\right]- \nonumber\\
&-\frac{\hbar^2}{8}(\dot{L}(0)L(0))^2\left[\delta_{n'\ell}
J_3(n,\ell)+ \delta_{n\ell} J_3(n',\ell)\right.\nonumber\\
&\left.+2J_2(n',\ell)J_2(n,\ell) \right].
\end{align}
Therefore $I_1$ is expressed as
\begin{align}
I_1 &\sim -\frac{1}{2}-\frac{i\pi\hbar}{2} \dot{L}(0)L(0) \nonumber \\
&\times \left[\sum_n n
J_1(n,\ell)J_2(n,\ell)-\ell
\sum_{n'} J_1(\ell,n')J_2(n',\ell) \right].
\end{align}
$J_1, J_2$ and $J_3$ used above are given respectively  by
\begin{align}
J_1(n,\ell)&=2\int_{0}^{1}y \sin(\ell\pi y)\cos(n\pi y)dy \nonumber \\
&=\left\{
\begin{array}{rcl}
-\frac{1}{2n\pi}, \quad $for$\quad n=\ell \\
(-1)^{n+\ell+1}\frac{2\ell}{(\ell^2-n^2)\pi} \quad $otherwise$,
\end{array}
\right.
\end{align}
\begin{align}
J_2(n,\ell)&=2\int_{0}^{1} y^2\sin(\ell\pi y)\sin(n\pi y)dy \nonumber \\
&=\left\{
\begin{array}{rcl}
\frac{1}{3}-\frac{1}{2n^2\pi^2}, \quad $for$\quad n=\ell \\
(-1)^{n+\ell}\frac{8n\ell}{(n^2-\ell^2)^2\pi^2} \quad $otherwise$
\end{array}
\right.
\end{align}
and
\begin{align}
J_3(n,\ell)&=2\int_{0}^{1} y^4\sin(\ell\pi y)\sin(n\pi y)dy \nonumber \\
&=\left\{
\begin{array}{rcl}
\frac{1}{5}-\frac{1}{n^2\pi^2}+\frac{3}{2 n^4\pi^4}, \\
\quad $for$\quad n=\ell \\
(-1)^{n+\ell}\left(\frac{16n\ell}{(n^2-\ell^2)^2\pi^2} - \frac{192n\ell(n^2+\ell^2)}{(n^2-\ell^2)^4\pi^4} \right)\\
\quad $otherwise$.
\end{array}
\right.
\end{align}


\begin{thebibliography}{0}
\bibitem{comm} The wave function in Section \ref{exact} describes a statistically pure state whereas the density operator in Section \ref{perturb} is concerned with a mixed state causing the correlation among initial states, which resulted in a discrepancy in the treatment of statistical weights.
At zero temperature and for the particle number $N \gg 1$, for example, this discrepancy amounts to $|C| \sim N$ and $|C^{\prime}| \sim N^3$, according to our numerical calculations.
\end{thebibliography}
\end{document}